\documentclass[conference]{IEEEtran}
\PassOptionsToPackage{draft}{hyperref}

\usepackage{xcolor}
\usepackage[linesnumbered,ruled,vlined]{algorithm2e}
\usepackage{bm}
\usepackage{amssymb}
\usepackage{comment}
\usepackage{color}

\usepackage{caption}
\usepackage{subcaption}

\ifCLASSINFOpdf
   \usepackage[pdftex]{graphicx}
  \graphicspath{ {./images/} }
   \DeclareGraphicsExtensions{.pdf,.jpeg,.png}
\else
\fi
\usepackage{amsmath}
\usepackage{mathrsfs}

\DeclareMathOperator*{\argmin}{arg\,min}

\usepackage{acro}
\providecommand\tooltip[2]{[PDF:no. of 1][tooltip:no. of 2]}
\acsetup{tooltip=true}

\DeclareAcronym{NN}{
  short        = {NN},
  long         = {neural network},
  tooltip      = {Neural Network}
}
\DeclareAcronym{MLP}{
  short        = {MLP},
  long         = {Multilayer Perceptron},
  tooltip      = {Multilayer Perceptron}
}

\DeclareAcronym{TIDAC}{
  short        = {TIDAC},
  long         = {time-interleaved DAC},
  tooltip      = {time-interleaved DAC}
}

\DeclareAcronym{DFO}{
  short        = {DFO},
  long         = {derivative-free optimization},
  tooltip      = {derivative-free optimization}
}

\DeclareAcronym{DSP}{
  short        = {DSP},
  long         = {digital signal processing},
  tooltip      = {digital signal processing}
}

\DeclareAcronym{ADC}{
  short        = {ADC},
  long         = {analog-to-digital converter},
  tooltip      = {analog-to-digital converter}
}

\DeclareAcronym{DAC}{
  short        = {DAC},
  long         = {digital-to-analog converter},
  tooltip      = {digital-to-analog converter}
}

\DeclareAcronym{SPI}{
  short        = {SPI},
  long         = {Serial Peripheral Interface},
  tooltip      = {Serial Peripheral Interface}
}

\DeclareAcronym{RF}{
  short        = {RF},
  long         = {radio-frequency},
  tooltip      = {radio-frequency}
}

\DeclareAcronym{SFDR}{
  short        = {SFDR},
  long         = {spurious-free dynamic range},
  tooltip      = {spurious-free dynamic range}
}

\DeclareAcronym{RZ}{
  short        = {RZ},
  long         = {return-to-zero},
  tooltip      = {return-to-zero}
}

\begin{document}
%
\title{Machine Learning Based Image Calibration for a Twofold Time-Interleaved High Speed DAC}


\author{
    \IEEEauthorblockN{
        Daniel Beauchamp\IEEEauthorrefmark{1}\IEEEauthorrefmark{2} and
        Keith M. Chugg\IEEEauthorrefmark{2}}
    \IEEEauthorblockA{
        \begin{tabular}{cc}
            \begin{tabular}{@{}c@{}}
                \IEEEauthorrefmark{1}
                    Jariet Technologies, 103 W Torrance Blvd, Redondo Beach, CA 90277 \\
                \IEEEauthorrefmark{2}
                   Ming Hsieh Department of Electrical Engineering, \\
University of Southern California, Los Angeles, California 90089 \\
                    \{dbeaucha, chugg\}@usc.edu
            \end{tabular} 
        \end{tabular}
    }
}


%


\maketitle

\begin{abstract}
In this paper, we propose a novel image calibration algorithm for a twofold \ac{TIDAC}. The algorithm is based on simulated annealing, which is often used in the field of machine learning to solve \ac{DFO} problems. The \ac{DAC} under consideration is part of a digital transceiver core that contains a high speed \ac{ADC}, microcontroller, and digital control via a \ac{SPI}. These are used as tools for designing an algorithm which suppresses the interleave image to the noise floor. The algorithm is supported with experimental results in silicon on a 10-bit twofold \ac{TIDAC} operating at a sample rate of 50 GS/s in 14nm CMOS technology.
\end{abstract}


%
\IEEEpeerreviewmaketitle

\section{Introduction}
Conventional \ac{RF} front-ends are typically composed of several mixers, local oscillators and analog filters. These components are a sizeable expense in terms of cost, area, and power, especially when implemented in phased array systems with several radiating antenna elements \cite{phased_array_power}. Fortunately, integrated circuit technology has advanced to such a degree that conventional \ac{RF} front-end solutions are being replaced with high speed \ac{ADC}s, \ac{DAC}s and \ac{DSP} which perform frequency conversion and filtering operations in the digital domain \cite{phased_array_cmos}. This allows data converters to be placed closer to the antenna, thereby significantly reducing system cost and power consumption. In addition, high speed converters have their thermal and quantization noise power spread across a wide Nyquist zone, which enhances dynamic range after processing gain. In order for data converters to achieve multi-GS/s rates, it is common to time-interleave several low speed converters \cite{TI_converters}, \cite{20G_DAC}. The high speed of the \ac{TIDAC} coupled with the area efficiency inherent in 14nm CMOS presents an ideal use case for phased array systems such as next generation radar and 5G. However, the inevitable timing errors and mismatch among the low speed converter slices results in images, or spectral replicas, which corrupt the converter output spectrum. Therefore, image calibration schemes are often necessary in order to avoid considerable loss of dynamic range. 

The authors in \cite{20G_DAC} consider a 20 GS/s 6-bit \ac{DAC} with no calibration scheme in place. As a result, the \ac{SFDR} is limited to 40 dB at output frequencies near 9 GHz. The authors in \cite{delta_sigma} consider a twofold delta sigma \ac{TIDAC} operating at an aggregate sample rate of 10 GS/s. The clock duty cycle error is understood to be the limiting impairment regarding dynamic range, and calibration schemes are proposed. However, the recommended solution involves digital pre-filtering, which is essentially equivalent to increasing the \ac{DAC} resolution and tightening matching requirements. Although an analog post-correction scheme is proposed, an accurate measurement of clock duty cycle is required, and this proves to be increasingly challenging at higher sample rates. \par In \cite{12b_DAC}, the issue of the interleave image is recognized as a limiting factor in high speed \ac{TIDAC} performance. A self-calibration circuit is proposed, but it is only functional for sample rates below 200 MS/s. Calibration schemes above this rate are left as an opportunity for future research. \par The authors in \cite{Olieman} provide a duty-cycle calibration algorithm for a twofold \ac{TIDAC}, but assume that the sub-DAC slices are balanced in terms of gain. Practically, this is not a valid assumption for an RF \ac{DAC} in deep sub-micron processes. In fact, even minor mismatch in sub-DAC gain can exacerbate the interleave image, leading to major loss of dynamic range. This is shown in Section II. 

In this paper, we consider a 10-bit twofold \ac{TIDAC} with current steering architecture operating at an aggregate rate of 50 GS/s using two 25 GS/s sub-DAC slices in 14nm CMOS technology. The \ac{DAC} is part of a digital transceiver core from Jariet Technologies that contains an on-chip high speed \ac{ADC}, microcontroller, and digital control via an SPI interface. For the \ac{DAC} under consideration, there is an image which appears at half of the aggregate sample rate. As far as we know, calibration schemes for \ac{DAC}s at sample rates this high have not been reported. As shown in Section II, the impairments which exacerbate this image are clock duty cycle error, mismatch in sub-DAC analog gain, and clock and data misalignment.

\begin{figure}[h]
\centering
\includegraphics[scale=0.4]{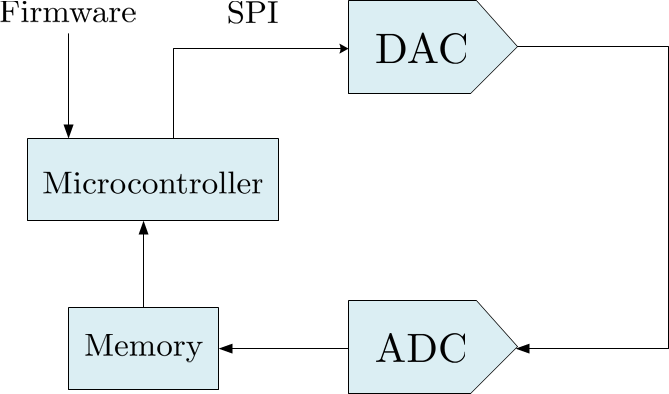}
\caption{Block diagram of the \ac{TIDAC} in a closed loop configuration.}
\label{fig:closed_loop}
\end{figure}

We use the closed-loop configuration shown in Fig. \ref{fig:closed_loop} to design an algorithm which suppresses the interleave image to the noise floor. This ensures that dynamic range does not suffer due to interleaving effects. Note that although the authors in \cite{Olieman} use a similar configuration to Fig. \ref{fig:closed_loop}, the algorithm proposed herein does not assume the sub-DACs are balanced in terms of gain. In addition, the configuration in Fig. \ref{fig:closed_loop} does not rely on any bandwidth limited circuitry as in \cite{12b_DAC}, and does not tighten matching requirements as in \cite{delta_sigma}. 

In Section II, we provide some background information on twofold \ac{TIDAC}s. Using Fourier analysis, we explicitly show how specific impairments can cause an undesired image at half of the aggregate sample rate. In Section III, we concretely define the problem at hand in an integer programming framework, and a novel solution is proposed based on simulated annealing. In Section IV, we apply this solution to a 50 GS/s \ac{DAC} in 14nm CMOS and provide experimental results which highlight its efficacy in terms of image suppression. We conclude in Section V by summarizing the key results and providing some direction for future research.

\section{Twofold time-interleaved DAC}
\begin{figure}[h]
\centering
\includegraphics[scale=0.4]{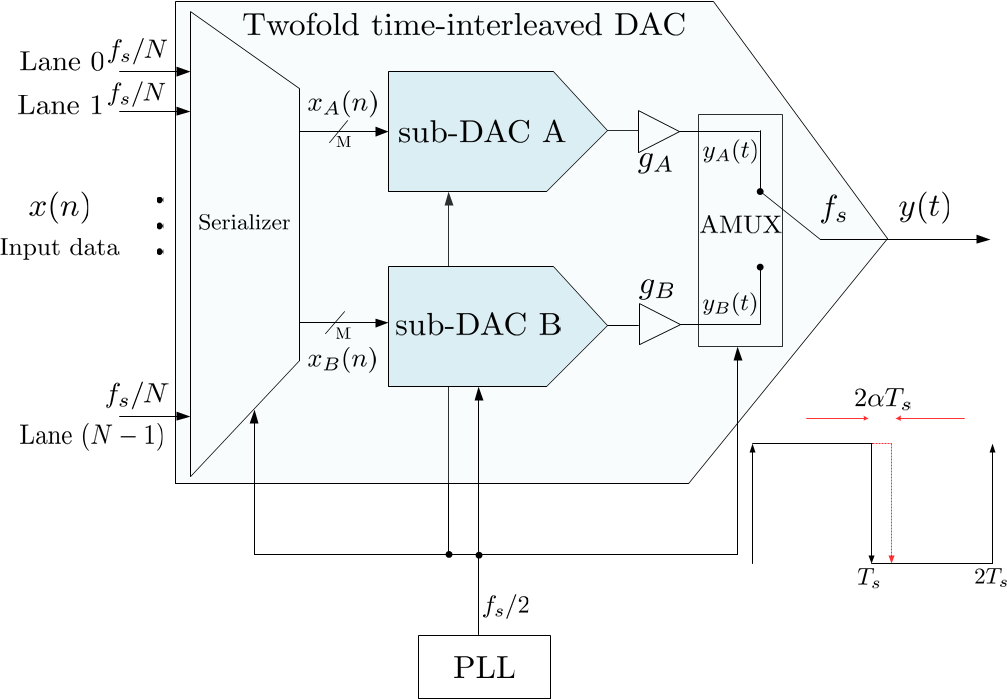}
\caption{Simplified block diagram of the \ac{TIDAC}, the fractional timing error, $\alpha$, shown in lower right.}
\label{fig:blockDiag}
\end{figure}

The block diagram for the general $M$-bit \ac{TIDAC} operating at a sample rate of $f_s$ is illustrated in Fig. \ref{fig:blockDiag}.

A phase locked loop (PLL) generates a clock at frequency $f_s/2$ which is distributed to the blocks denoted by serializer, sub-DAC A, sub-DAC B, and AMUX. The serializer contains a clock tree with several 2-to-1 multiplexers that serialize the $N$ low speed parallel lanes into two high speed ones at the $f_s/2$ rate. The sub-DAC slices employ current drivers for each bit to convert the $M$-bit code presented at the input to an analog output current. The drivers are composed of binary weighted current sources and clock driven switches. When the switches are active, current is driven to the output, and when they are inactive, current is dumped to a dummy node which is not shown in the diagram. This is controlled by the analog multiplexer (AMUX). \par Ideally, in this ping-pong like configuration, each sub-DAC drives current to the output for $50\%$ of the half-rate clock period. However, this is generally not the case due to unavoidable clock duty cycle error. In Fig. \ref{fig:blockDiag}, we include a fractional timing offset factor $\alpha \in [-1,1]$ in order to account for this. Note that $\alpha = 0$ corresponds to the ideal case of $50\%$ duty cycle. In this section, we show that this impairment causes an image in the frequency domain which is located at $f_s/2$. Also shown in Fig. \ref{fig:blockDiag} are the sub-DAC analog gains, $g_A$ and $g_B$. Note that in general, $g_A \neq g_B$ mainly due to current source imbalance between the sub-DACs, and this also causes an image at $f_s/2$. We refer to the architecture illustrated in Fig. \ref{fig:blockDiag} as a current steering twofold \ac{TIDAC}. We proceed by computing $Y(f)$, which is the Fourier transform of the \ac{DAC} output $y(t)$. Throughout the paper, we denote the Fourier transform of a time-domain signal $y(t)$ by
\begin{align} 
    Y(f) = \int_{-\infty}^\infty y(t)\ e^{-j 2 \pi f t}\, dt. \label{eq1}
\end{align}
Note that 
\begin{align}
y(t) = y_A(t) + y_B(t),
\label{eq2}
\end{align}
so we can compute the Fourier transform of the individual sub-DACs and then simply add the result to obtain $Y(f)$ by linearity of the Fourier transform. Without loss of generality, assume that sub-DAC A is driving current to the output at time $t=0$. Note that $y_A(t)$ can be modeled as a sum of phase shifted \ac{RZ} pulses whose amplitude is determined by the discrete-time sequence $x_A(n) = x(2nT_s)$, where $x(t)$ is the continuous-time representation of the input. In particular, we have
\begin{align}
y_A(t) &= g_A\ \Pi \left(\frac{t-\frac{T_s}{2}(1+2\alpha)}{T_s\left(1+2\alpha\right)}\right) \nonumber \\ &~~~~~ \ast \left( x(t) \cdot \sum_{k=-\infty}^{\infty} \delta\left(t - 2kT_s \right) \right)
\label{eq3}
\end{align}
 where
\begin{align}
    \Pi(t) := \begin{cases}
0 & \mbox{if } |t| > \frac{1}{2} \\
1 & \mbox{if } |t| = \frac{1}{2} \\
1 & \mbox{if } |t| < \frac{1}{2} \\
\end{cases}
\end{align}
$\delta(t)$ is the Dirac delta function, and $\ast$ denotes the convolution operator. It is clear that (\ref{eq3}) is a sum of phase shifted \ac{RZ} pulses, as it is the convolution of a rectangular function with an impulse train. Taking the Fourier transform of (\ref{eq3}), we have
\begin{align}
    Y_A(f) &= \frac{g_A}{2} \left(1+2 \alpha\right) \text{sinc} \left(f T_s \left(1+2\alpha\right) \right)\ e^{-j \pi f T_s \left(1+2\alpha\right)}  \nonumber \\ &~~~~~ \times  \sum_{k=-\infty}^{\infty} X\left(f-k\frac{f_s}{2}\right)
    \label{eq5}
\end{align}
where $\text{sinc}(x) :=\sin(\pi x)/(\pi x)$, and we use the fact that convolution in the time domain becomes multiplication in the frequency domain and vice-versa. The Fourier transform of $y_B(t)$ is obtained similarly, and is given by
\begin{align}
    Y_B(f) &= \frac{g_B}{2} \left(1-2 \alpha\right) \text{sinc} \left(f T_s \left(1-2\alpha\right) \right)\ e^{-j \pi f T_s \left(1-2\alpha\right)}  \nonumber \\  &~~~~~ \times \sum_{k=-\infty}^{\infty} X\left(f-k\frac{f_s}{2}\right) e^{-j \pi k \left(1+2\alpha\right)}
    \label{eq6}
\end{align}
Note the additional complex exponential factor in the sum of \eqref{eq6} compared to \eqref{eq5} due to the assumption that sub-DAC~A is aligned at $t=0$. Using (\ref{eq2}), the Fourier transform of the \ac{DAC} output $y(t)$ is
\begin{align}
Y(f) = Y_A(f) + Y_B(f)
\label{eq7}
\end{align}
where $Y_A(f)$ and $Y_B(f)$ are given by \eqref{eq5} and \eqref{eq6} respectively. Note that if $\alpha=0$, the complex exponential in the sum of \eqref{eq6} is -1 for $k$ odd and will cancel the the corresponding term in \eqref{eq5} if and only if $g_A = g_B$. As mentioned in the introduction, clock and data misalignment also exacerbates the $f_s/2$ image. 

\begin{figure}[h]
\centering
\includegraphics[scale=0.5]{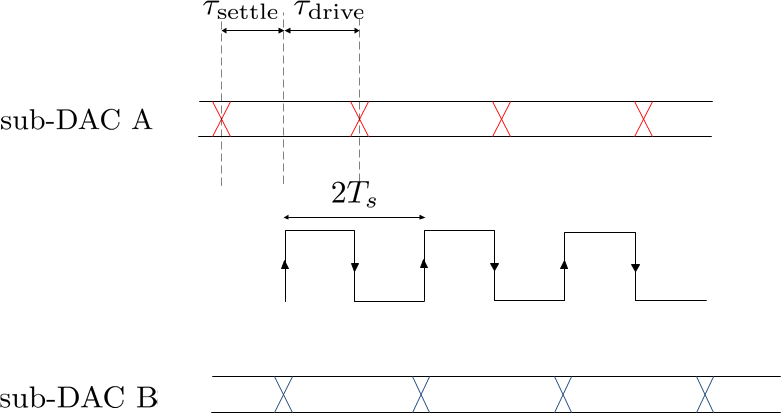}
\caption{Illustration of clock and data alignment for a twofold \ac{TIDAC}.}
\label{fig:blockDiag_2}
\end{figure}

When sub-DAC A undergoes a data transition, there is a settling window of $\tau_{\text{settle}}$ as shown in Fig. \ref{fig:blockDiag_2}. During this time, sub-DAC A is dumping current to the dummy node while sub-DAC B is driving current to the output. The ideal scenario corresponds to the case where the clock edges are equidistant from the data transitions as illustrated in Fig. \ref{fig:blockDiag_2}. In any other scenario, one sub-DAC has a longer (or shorter) $\tau_{\text{drive}}$ than the other. It is this timing imbalance which exacerbates the image at $f_s/2$ in a manner similar to that of clock duty cycle error. For the chip under consideration in Section IV, there is an algorithm that performs coarse clock and data alignment, but that is beyond the scope of this paper. \par
\begin{figure}[ht]
\centering
\includegraphics[scale=0.6]{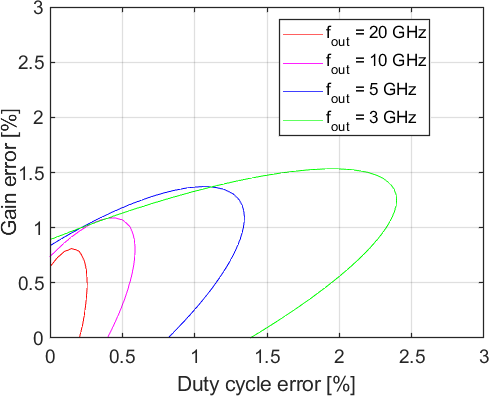}
\caption{-50 dBc level curves of interleave image magnitude when the ideal output is a sinusoid at frequency $f_{\text{out}}$.}
\label{fig:fig4}
\end{figure}

 Consider a twofold \ac{TIDAC} for the case in which the ideal output is a sinusoid at frequency $f_{\text{out}}$. From inspection of \eqref{eq7}, there is an interleave spur which appears at $f_s/2 - f_{\text{out}}$. The contour plots in Fig. \ref{fig:fig4} illustrate the -50 dBc level curves of the interleave spur magnitude for various values of $f_{\text{out}}$. These are obtained using \eqref{eq7}. If the gain and duty cycle errors are contained within these contours on the lower left region of Fig. \ref{fig:fig4}, then we guarantee the image spur is less than -50 dBc, which is reasonable from an SFDR perspective for a wideband RF \ac{DAC}. From Fig. \ref{fig:fig4} it is clear that extremely small gain and duty cycle errors are required for reasonable \ac{DAC} \ac{SFDR} performance. 

In Section III, we propose a machine learning based algorithm which uses digital control to suppress the interleave spur to the noise floor. 

\section{Simulated Annealing Algorithm}
As mentioned in the introduction, the \ac{DAC} under consideration is part of a digital transceiver core that contains a high speed \ac{ADC} and digital control via a microcontroller and SPI interface. There are several controls which remedy the impairments discussed in Section II. Table \ref{tab:control_table} outlines these controls along with their corresponding objectives. Note that the chip under consideration in Section IV has these controls split into six different control registers, each of which has a wide range of discrete settings. Therefore, we begin by defining a state vector $\bm{s} \in \mathcal{S} \subset \mathbb{R}^6$  whose entries are composed of the digital control settings. In order to find the optimal control settings, we require the ability to measure the interleave spur power. Consider a \ac{TIDAC} with sample rate $f_s$ and sinusoidal output with frequency $f_{\text{out}}$. Again, by inspection of \eqref{eq7}, we observe that an interleave spur appears at $f_s/2-f_{\text{out}}$. Using the on-chip \ac{ADC}, we then sample the \ac{DAC} output, compute the fast Fourier Transform (FFT), and monitor the bin corresponding to $f_s/2 - f_{\text{out}}$. The energy in this FFT bin then defines a cost function $C:\mathcal{S} \rightarrow \mathbb{R}$. The objective is to then choose a vector $\bm{s}^* \in \mathcal{S}$ such that

\begin{align}
\bm{s}^* = \argmin_{\bm{s}} C(\bm{s})
\label{eq8}
\end{align}

\begin{table}[]
    \centering
    \begin{tabular}{|c|c|} \hline
    Digital control & Objective \\ \hline \hline
        sub-DAC output current  & $g_A \rightarrow g_B$  \\
            \hline $f_s/2$ clock duty cycle & $\alpha \rightarrow 0$
          \\  \hline Phase rotator  & $\tau_{\text{settle}} \approx  \tau_{\text{drive}}$ for both sub-DACs \\ \hline
    \end{tabular}
    \caption{List of digital controls with corresponding objectives.}
    \label{tab:control_table}
\end{table}

The objective defined by \eqref{eq8} is an integer programming problem. There are a couple of key items worth mentioning. First, note that we do not have an expression for the cost function $C(\bm{s})$, so optimization via relaxation and differentiation is not an option. In addition, the solution space is large, as the state vector lies in six-dimensional space and each entry has a wide range of discrete values. A suitable algorithm which promotes global optimum convergence in this scenario is known as simulated annealing \cite{dfo}. The pseudocode for simulated annealing is outlined in Algorithm \ref{alg:simulated_annealing}.

\SetKwInput{KwInput}{Input}           
\SetKwInput{KwOutput}{Output}
\newcommand\mycommfont[1]{\footnotesize\ttfamily\textcolor{blue}{#1}}
\SetCommentSty{mycommfont}

\begin{algorithm}
\DontPrintSemicolon
  
  \KwInput{$\bm{s}_0, T_{\text{max}}, T_{\text{min}}, \gamma, \beta$, $K$}
  \KwOutput{$\bm{s}^*$} \BlankLine
  $\bm{s} \gets \bm{s_0}$ \;
  $\bm{s}^* \gets \bm{s}$ \;
  $T \gets T_{\text{max}}$ \;
  \While{$T>T_{\text{min}}$}
  {
  \For{$k \gets 0$ to $K-1$}
  {
  $\bm{s}^\prime \gets n(\bm{s})$ \;
  $\Delta E \gets C(\bm{s}^\prime) - C(\bm{s})$ \;
  \If{$\Delta E \leq 0$}
    {
        $\bm{s} \gets \bm{s}^\prime$\;
        \If{$C(\bm{s}) < C(\bm{s}^*)$}
        {
        $\bm{s}^* \gets \bm{s}$\;
        }
    }
    \ElseIf{$\text{rand}(0,1) < \exp {\left(-\beta \frac{\Delta E}{T} \right)}$}
    {
        $\bm{s} \gets \bm{s}^\prime$\;
    }
  }

    $T \gets \gamma T$
 }

\caption{Simulated annealing.}
\label{alg:simulated_annealing}
\end{algorithm}

Algorithm \ref{alg:simulated_annealing} has a temperature parameter $T$ which starts high at $T_\text{max}$ and gradually reduces to $T_\text{min}$ exponentially with factor $\gamma$. At each value of $T$, we perform $K$ iterations which involve a cost comparison of the current state $\bm{s}$ with a neighboring state $\bm{s}^\prime = n(\bm{s})$. Note that states $\bm{s}^\prime$  whose cost is less than or equal to the current state $\bm{s}$ are always accepted (i.e. $\Delta E \leq 0$). If a neighbor is accepted under the criteria $\Delta E \leq 0$, then we check whether or not it has a lower cost than the optimal state $\bm{s}^*$. However, states with higher cost (i.e. $\Delta E > 0$) are not necessarily rejected. In fact, the acceptance of higher cost states is controlled by the temperature $T$ in a probabilistic manner. Note that the term $\exp\left( {-\beta \frac{\Delta E}{T}}\right) \rightarrow 1$ as $T \rightarrow \infty$ where $\beta>0$ is a hyperparameter. This implies that the state space is explored aggressively when $T$ is large since the acceptance of higher cost states becomes more probable. A key component of Algorithm \ref{alg:simulated_annealing} involves constructing the neighboring state function $n(\bm{s})$. In our case, this process first involves choosing a number from the discrete uniform distribution $\mathcal{U}\left\{1,6\right\}$ which corresponds to one of the six digital controls. We then choose another number uniformly at random over a range which covers the selected control setting. The neighbor state is found by simply substituting the new control setting into a copy of the previous state. 

\section{Experimental Results}
In this section, we use the Agilent N9030A spectrum analyzer to apply Algorithm \ref{alg:simulated_annealing} to a 10-bit twofold 50 GS/s \ac{TIDAC} in 14nm CMOS. Note that the spectrum analyzer samples the DAC output which effectively emulates the on-chip ADC.

\begin{figure}[h]
\centering
\includegraphics[scale=0.6]{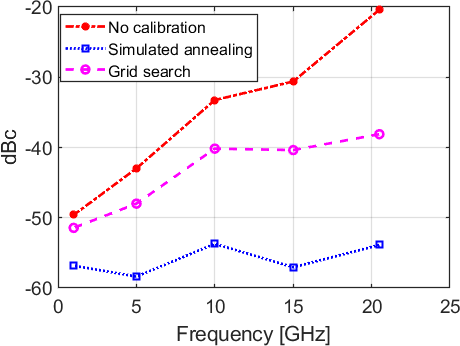}
\caption{Interleave spur performance over Nyquist.}
\label{fig:pre_vs_post_cal}
\end{figure}
\begin{figure}[ht]
\centering
\includegraphics[scale=0.07]{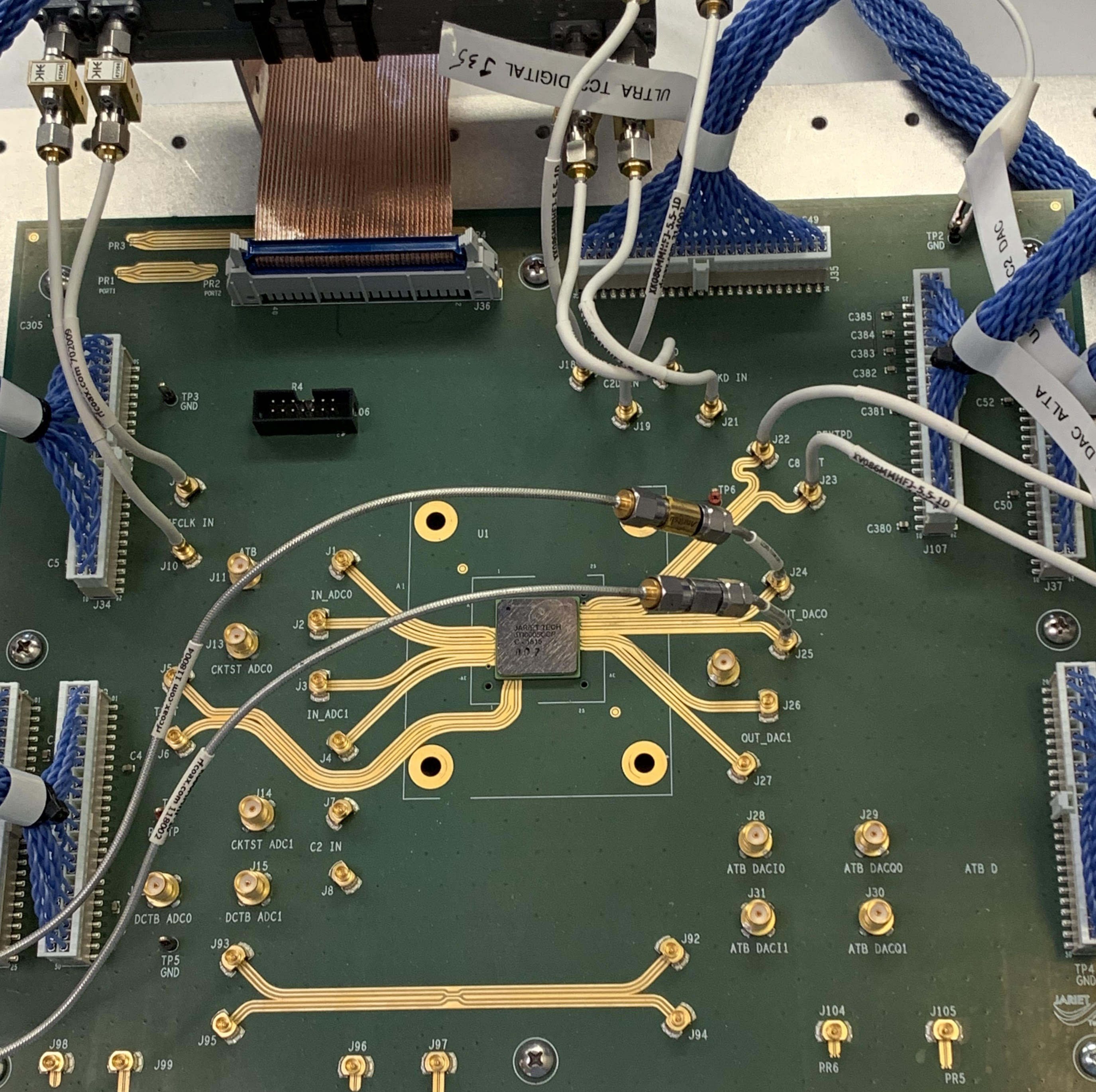}
\caption{Test board with digital transceiver chip containing a 50 GS/s \ac{TIDAC}.}
\label{fig:test_setup}
\end{figure}

The plot in Fig. \ref{fig:pre_vs_post_cal} demonstrates the efficacy of Algorithm \ref{alg:simulated_annealing} over Nyquist and compares it to a simple grid search over the state space. Note that Algorithm \ref{alg:simulated_annealing} keeps the interleave spur well below -50 dBc.  After starting Algorithm \ref{alg:simulated_annealing} with control registers in their initial states, convergence occurs after an average of 160 interleave spur measurements, and grid search was performed with 280 measurements. Note that at high frequency, simulated annealing has a 15 dB improvement over grid search while requiring nearly half as many measurements. The parameters used as input to Algorithm \ref{alg:simulated_annealing} were $\gamma = 0.8$, $K = 30$, and $\beta = 50$. These experiments were conducted using an Altera FPGA which serves as a bridge between the PC and the SPI interface. The test board and chip are shown in Fig. \ref{fig:test_setup}.

\section{Conclusion}
In this paper, a novel image calibration algorithm for a twofold \ac{TIDAC} is proposed and verified in silicon on a 10-bit 50 GS/s \ac{DAC} in 14nm CMOS. The algorithm does not exacerbate matching requirements as in \cite{delta_sigma}, and does not assume the sub-DAC gains are balanced as in \cite{Olieman}. Furthermore, bandwidth limited calibration circuitry is not required as in \cite{12b_DAC}. Although an on-chip high speed \ac{ADC} is assumed, this is becoming much more practical with the use of low power deep sub-micron processes like 14nm CMOS. Future work involves repeating the measurements in Section IV using the on-chip \ac{DAC} to \ac{ADC} loopback path. Beyond interleave impairments, high speed data converters have harmonic distortion. Using machine learning for harmonic suppression would be another interesting and fruitful research opportunity.  

\section*{Acknowledgment}
The authors would like to Jariet Technologies for providing the financial support and equipment which made this research possible.



%


\bibliographystyle{IEEEtran}
\bibliography{}


\end{document}